\documentclass[twocolumn, prX]{revtex4-1} 

\usepackage[ansinew]{inputenc}
\usepackage[T1]{fontenc}
\usepackage{makeidx,amsfonts,amssymb,amsmath}
\usepackage{graphicx}
\usepackage{epstopdf}
\usepackage{color}
\usepackage[caption=false]{subfig}		
\usepackage[abs]{overpic} 
\usepackage[rightcaption]{sidecap} 
\sidecaptionvpos{figure}{h} 

\newcommand{\enm}{\mbox{$\;$nm}}    	
\newcommand{\emum}{\mbox{$\;$$\mu$m}}    	
\newcommand{\eK}{\mbox{$\;$K}}			
\newcommand{\eA}{\mbox{$\;$A}}			
\newcommand{\eW}{\mbox{$\;$W}}			
\newcommand{\eProzent}{\mbox{$\;$\%}}    	

\begin{document}

\title{Transport and Capture Properties of Auger-Generated High-Energy Carriers in (AlInGa)N Quantum Well Structures} 

\author{A. Nirschl}
\email{Anna.Nirschl@osram-os.com}
\affiliation{OSRAM Opto Semiconductors GmbH, Leibnizstraße 4, 93055 Regensburg, Germany}
\affiliation{Institut für Experimentelle und Angewandte Physik, Universität Regensburg, Universitätsstraße 31, 93040 Regensburg, Germany}

\author{M. Binder}
\affiliation{OSRAM Opto Semiconductors GmbH, Leibnizstraße 4, 93055 Regensburg, Germany}

\author{M. Schmid}
\affiliation{OSRAM Opto Semiconductors GmbH, Leibnizstraße 4, 93055 Regensburg, Germany}

\author{M. M. Karow}
\thanks{Present address: Institut für Festkörperphysik, Technische Universität Berlin, Hardenbergstraße 36, 10623 Berlin, Germany}
\affiliation{OSRAM Opto Semiconductors GmbH, Leibnizstraße 4, 93055 Regensburg, Germany}

\author{I. Pietzonka}
\affiliation{OSRAM Opto Semiconductors GmbH, Leibnizstraße 4, 93055 Regensburg, Germany}

\author{H.-J. Lugauer}
\affiliation{OSRAM Opto Semiconductors GmbH, Leibnizstraße 4, 93055 Regensburg, Germany}

\author{R. Zeisel}
\affiliation{OSRAM Opto Semiconductors GmbH, Leibnizstraße 4, 93055 Regensburg, Germany}

\author{M. Sabathil}
\affiliation{OSRAM Opto Semiconductors GmbH, Leibnizstraße 4, 93055 Regensburg, Germany}

\author{D. Bougeard}
\affiliation{Institut für Experimentelle und Angewandte Physik, Universität Regensburg, Universitätsstraße 31, 93040 Regensburg, Germany}

\author{B. Galler}
\affiliation{OSRAM Opto Semiconductors GmbH, Leibnizstraße 4, 93055 Regensburg, Germany}


\begin{abstract}
Recent photoluminescence experiments presented by M. Binder \textit{et al.} [Appl. Phys. Lett. $\textbf{103}$, $071108$ ($2013$)] demonstrated the visualization of high-energy carriers generated by Auger recombination in (AlInGa)N multi quantum wells. Two fundamental limitations were deduced which reduce the detection efficiency of Auger processes contributing to the reduction in internal quantum efficiency: the capture probability of these hot electrons and holes in a detection well and the asymmetry in type of Auger recombination. We investigate the transport and capture properties of these high-energy carriers regarding polarization fields, the capture distance to the generating well and the capture volume. All three factors are shown to have a noticeable impact on the detection of these hot particles. Furthermore, the investigations support the finding that electron-electron-hole exceeds electron-hole-hole Auger recombination if the densities of both carrier types are similar. Overall, the results add to the evidence that Auger processes play an important role in the reduction of efficiency in (AlInGa)N based LEDs. 
\end{abstract}

\pacs{}

\maketitle 
\section{Introduction}
After years of continuously improving the efficiency of GaN-based light-emitting diodes (LEDs), some fundamental properties are still lacking understanding. In particular, the so-called droop phenomenon, the reduction of the internal quantum efficiency (IQE) with increasing current density, \cite{Mukai1999, Piprek2010, Schubert2013} is discussed controversially in literature. \cite{Hangleiter2005, Shen2007, Schubert2007, David2010, LaubschIEEE} Recently, two sophisticated experimental approaches yielded direct evidence for the important role of Auger recombination by detecting high-energy carriers at a rate proportional to the droop. \cite{ Iveland2013, Binder2013} 
Iveland et al. \cite{Iveland2013} performed electron emission spectroscopy on a cesiated InGaN/GaN LED under electrical injection. While increasing the current, droop was observed along with an increase of the detected high-energy electrons which were attributed to Auger-processes in the quantum wells (QWs). The authors showed that the missing current due to droop directly correlates to the measured Auger electron current with a ratio of $10^{-6}$ -- a strong argument that the reduction in efficiency partly results from electron-electron-hole Auger recombination.
Binder et al. \cite{Binder2013} measured photoluminescence on a test structure alternatingly containing QWs emitting ultra-violet (UV) and green luminescence, respectively. Upon excitation at a wavelength that can generate carriers solely in the green wells, up to carrier densities at which the emission efficiency showed significant droop, UV luminescence was observed. The authors showed that these high-energy carriers partly account for the green photons missing due to droop and calculated a lower limit of $1\eProzent$ of the droop which must be ascribed to Auger processes. 

Hence, a direct proportionality of the droop-induced missing current to the measured high-energy carriers, which arise from Auger recombination, was observed in two different experimental approaches.
Although both findings indicate the importance of Auger recombination for the droop phenomenon, the obtained detection efficiencies are still low due to fundamental limitations of each measurement setup and the underlying physics. Indeed, the former experiment is only sensitive to electron-electron-hole Auger processes and the latter is limited by the need for both types of high-energy carriers to recombine radiatively. 
Both groups plausibly argue that the actual percentage of the contribution of Auger recombination to the droop must be much higher, but a fully quantitative demonstration is still pending.

In this article, we examine one of the limitations decreasing the detection efficiency in the approach by Binder et al. \cite{Binder2013}. High-energy carriers, stemming from Auger recombination, need to be captured by the UV QWs to be visualized. Hence, any capture probability below $100\eProzent$ reduces the fraction of high-energy carriers which is detectable. The following three factors are shown to influence the capture probability: the polarization fields, the distance between generation and capture well, and the capture volume.
\begin{figure*}[t!]
	\begin{minipage}[t]{0.28\textwidth}
	\textbf{(a)}	\textcolor{blue}{\textbf{g-uv}}\\
	\centering
		\includegraphics[width=0.85\textwidth,height=4.8cm]{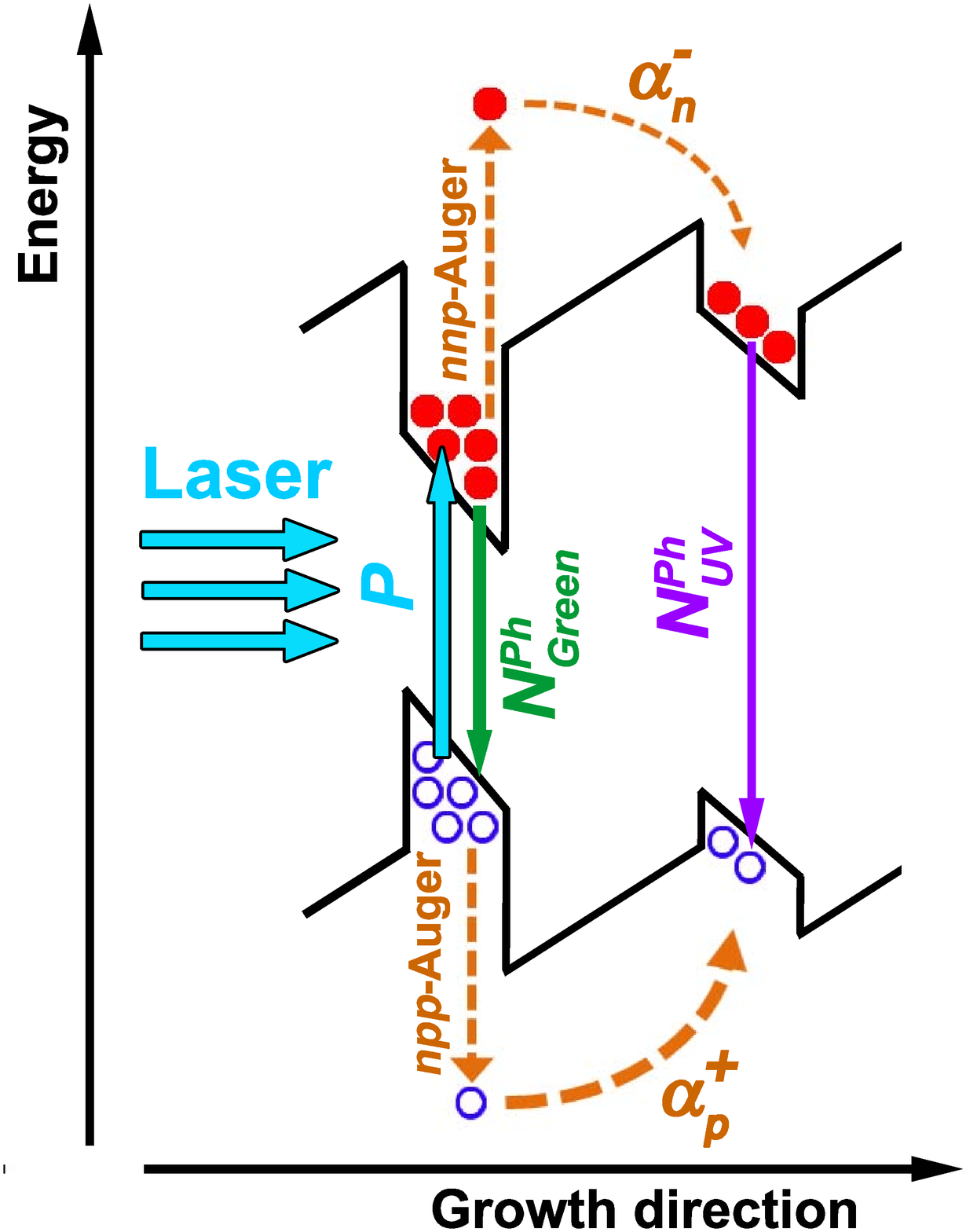} 
\par\vspace{0.3cm} 
				\large{$\beta_{g-uv}=\frac{C_{nnp}+C_{npp}}{\text{min}[\alpha^-_{n}\cdot C_{nnp},\alpha^+_{p}\cdot C_{npp}]}$}
			\end{minipage} 
				\begin{minipage}[t]{0.28\textwidth}
					\textbf{(b)} \textcolor[rgb]{0,0.58,0}{\textbf{uv-g}}\\
				\centering
		\includegraphics[width=0.9\textwidth,height=4.8cm]{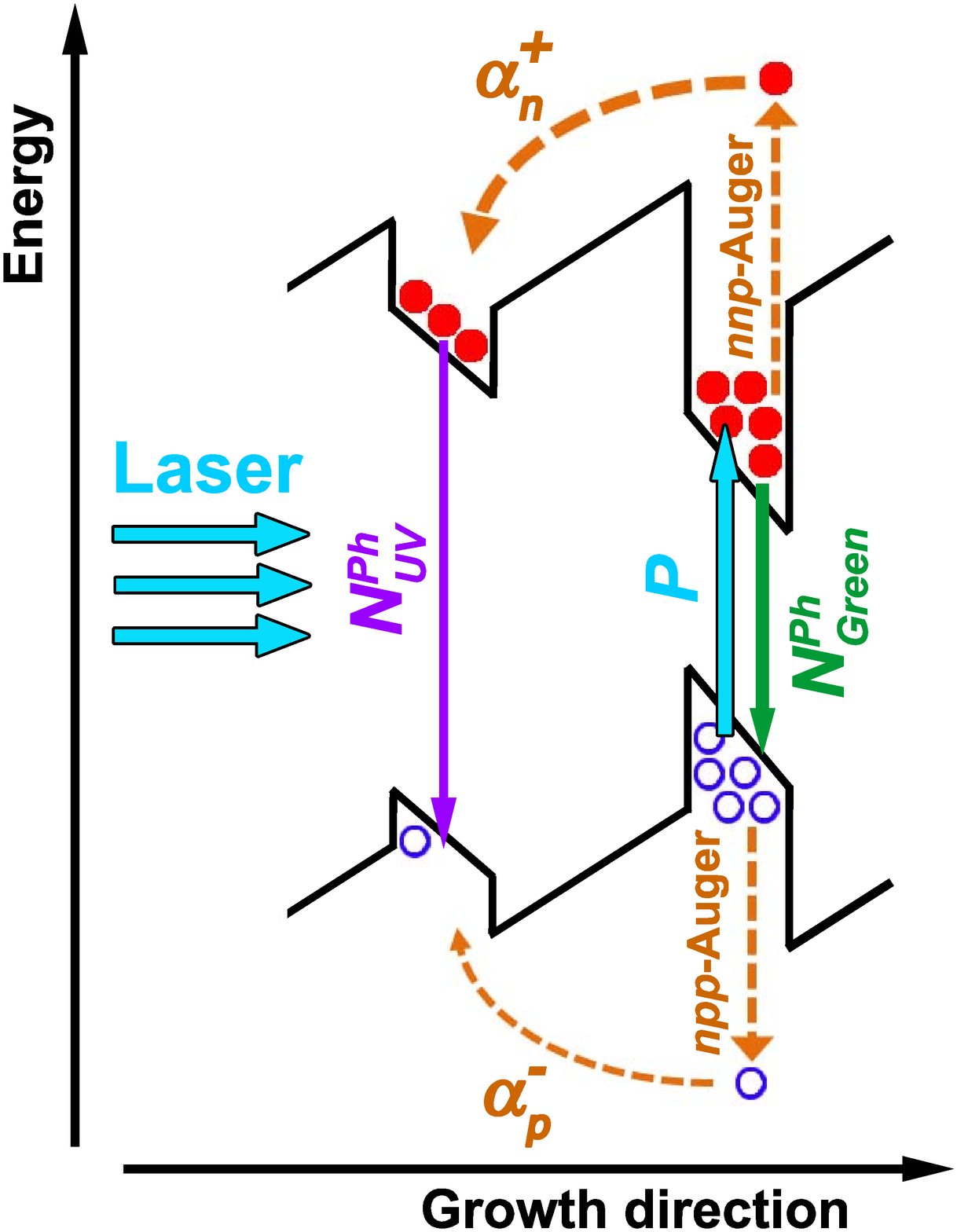} 
		\par\vspace{0.3cm} 
				\large{$\beta_{uv-g}=\frac{C_{nnp}+C_{npp}}{\text{min}[\alpha^+_{n}\cdot C_{nnp},\alpha^-_{p}\cdot C_{npp}]}$}
	\end{minipage}
					\begin{minipage}[t]{0.41\textwidth}
\textbf{(c)} \textcolor{red}{\textbf{g-uv-g}}\\
				\centering
		\includegraphics[width=0.8\textwidth,height=4.8cm]{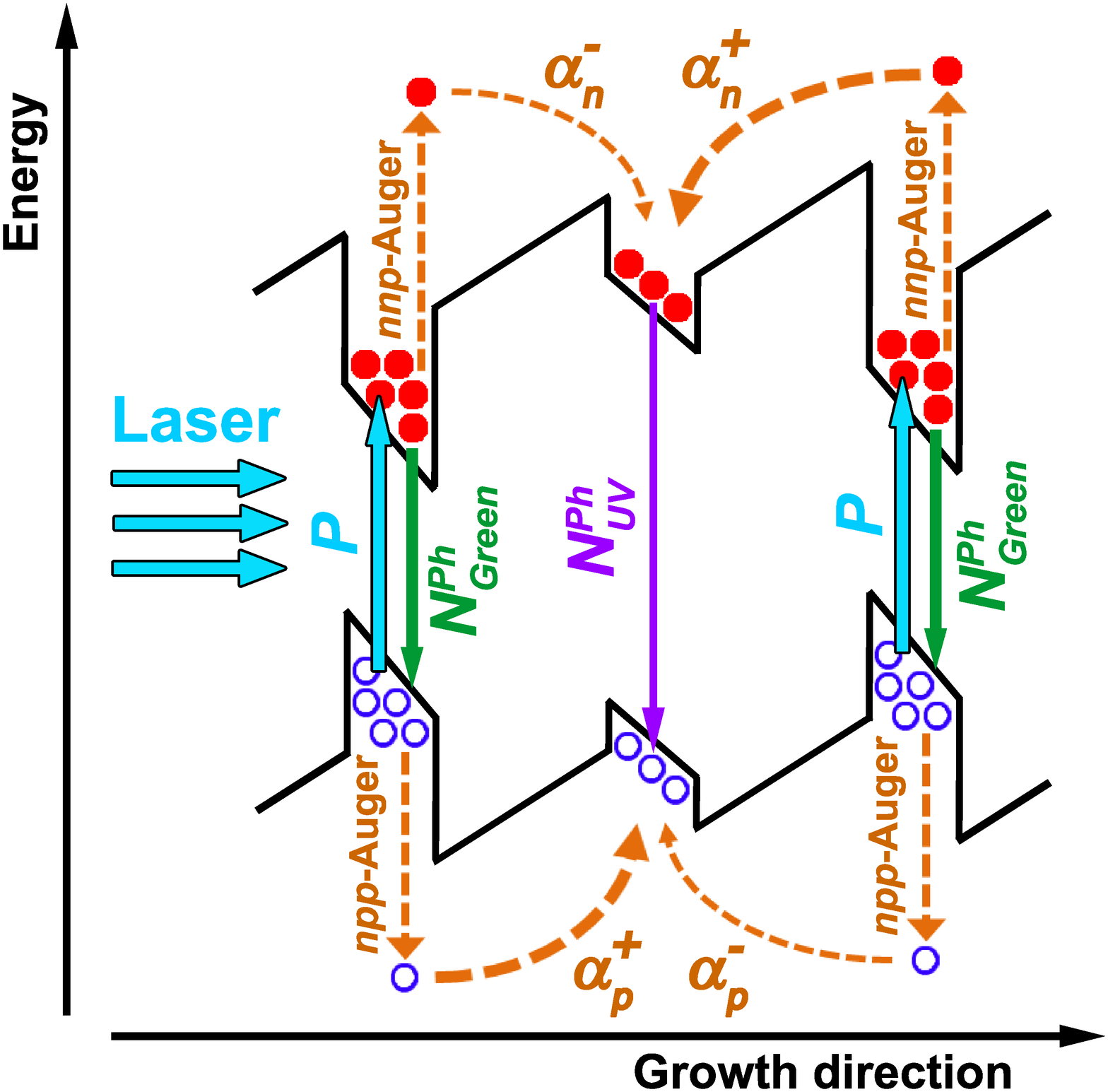}
		\par\vspace{0.3cm} 
		\large{$\beta_{g-uv-g}=\frac{2\cdot(C_{nnp}+C_{npp})}{\text{min}[(\alpha^-_{n}+\alpha^+_{n})\cdot C_{nnp},(\alpha^-_{p}+\alpha^+_{p})\cdot C_{npp}]}$}
	\end{minipage}
	\caption{\label{fig:Bandstruktur} Schematic illustration of the bandstructures discussed in section \ref{ssec: Pol} under the influence of polarization fields. The possible influence of the electric fields on the capture probability of high-energy electrons and holes is depicted. }
\end{figure*}

\section{Method}
All samples discussed in this article were grown by metal-organic vapor phase epitaxy on ($0001$) sapphire substrates. The deposition started with a low-temperature GaN nucleation layer, followed by a $4\emum$ thick undoped GaN buffer. While the structures differ in the particular design of the active region, they all comprise $3\enm$ thick ultra-violet and green emitting InGaN QWs with $10\eProzent$ and $22\eProzent$ indium content, respectively. Unless stated otherwise, they are separated by $7\enm$ thick Al$_{0.39}$Ga$_{0.61}$N barriers. Finally, the epitaxial stacks were capped with a $10\enm$ GaN layer.
The active regions were adapted to investigate different influences on the capture probability of high-energy carriers. Therefore, the arrangement and the number of detection and source well of Auger-generated carriers were varied as described in detail in the respective section.

Photoluminescence (PL) measurements were carried out at $12\eK$ using a $450\enm$ laser diode with an output power of $1.2\eW$ at a driving current of $1\eA$. This setup generates carrier densities in the green QWs at which the emission efficiency already drops, but does not directly excite electron-hole-pairs in the UV wells. The observation of UV luminescence is then a direct proof of both high-energy electrons and holes being generated in Auger processes taking place in the green QWs and being captured in the UV QWs. Other proposed explanations for the occurrence of UV luminescence could be excluded in previous work. \cite{Binder2013}
From these considerations, it is obvious that the detection probability of Auger-type efficiency losses is limited by the smaller density of the two types of carriers captured in the UV QWs. Any asymmetry between the electron-electron-hole ($nnp$-) and electron-hole-hole ($npp$-) Auger recombination rate or capture probability of the two types of carriers below $100\eProzent$ reduces the fraction of Auger processes which can be visualized.

To quantitatively evaluate the contribution of Auger recombination to the droop, we follow our previous approach \cite{Binder2013} where we determine the factor between the green photons which are lost due to droop and the measured UV photons. A measure for the IQE can be obtained from the number of emitted green photons divided by the excitation power. Due to Shockley-Read-Hall (SRH) recombination freezing out, \cite{Shockley1952a} the absolute IQE remains constant in the low excitation regime and can thus be considered to be $100\eProzent$. The missing green photons in arbitrary units can be determined from the difference of this plateau to the reduced efficiency caused by the droop.
Next, the lost photons are expressed via the measured UV photons multiplied by a constant $\beta$. This can be done since both the number of UV photons as well as the droop have been shown to scale with the cubed carrier density in the green QW. \cite{Binder2013} 
Using the same arbitrary units for the number of UV photons as for the calculation of the missing green photons, $\beta$ can be calculated for every excitation density $P$ and must be constant for a given sample:
\begin{equation}
\beta:=\frac{\text{max}\left[ \frac{N^{Ph}_{Green}}{P}\right]-\frac{N^{Ph}_{Green}}{P}}    {\frac{N^{Ph}_{UV}}{P}},
\label{eq: calc}
\end{equation} 
where $N^{Ph}_{Green}$ and $N^{Ph}_{UV}$ are the number of green and UV photons, respectively. 
Every measurement was repeated several times to minimize uncertainty due to statistical error. 
If Auger recombination is the only origin of the droop, $2/\beta$ is the fraction of the total Auger recombination which can be visualized as UV luminescence:
\begin{equation}
\beta=\frac{C_{nnp}+C_{npp}}{\text{min}[\alpha_{n}\cdot C_{nnp},\alpha_{p}\cdot C_{npp}]}, 
\label{eq: beta}
\end{equation} 
where $C_{nnp}$ and $C_{npp}$ are the coefficients for $nnp$- and $npp$-Auger processes and $\alpha_{n}$ and $\alpha_{p}$ are the capture probabilities for high-energy electrons or holes in the UV QWs. Consequently, the proportionality factor $\beta$ is limited by the possible type asymmetry in strength of Auger recombination and by the fraction of high-energy carriers which are not captured by the UV QW. In the following, three factors influencing the capture probability are discussed.

\section{results and discussion}
\subsection{Directionality due to Polarization Fields}
\label{ssec: Pol}
First, the influence of the polarization fields on the transport of high-energy carriers is evaluated. For this purpose, the position of the generation and capture wells within the layer stack are changed with respect to the polarization fields. The schematic bandstructures of the samples are depicted in Fig. \ref{fig:Bandstruktur}. The labeling follows the order of the QW type in growth direction. Since the generation of high-energy carriers in all structures results from Auger recombination in green QWs grown identically, the Auger coefficients are assumed to be equal. Apparently, the detection of high energy carriers generated by the green QWs in the structures g-uv and uv-g would be the same and thus 
\begin{equation}
\beta_{g-uv}=\beta_{uv-g}
\label{eq: PolNoInfluence}
\end{equation}
would hold, if the polarization fields and thus the arrangement of wells do not have any influence on the capture probabilities.

However, if the polarization fields have an influence on the transport of the high-energy carriers, electrons which are scattered into the barriers will preferably be driven into the direction of the substrate, when considering for example a simple description based on the "charged particle in an electric field model". 
In contrast, holes will be accelerated into the opposite direction. For the structure g-uv in Fig. \ref{fig:Bandstruktur}(a), this would lead to a lower capture probability for electrons $\alpha^-_{n}$ and to a higher one for holes $\alpha^+_{p}$. 
Comparing structure uv-g in Fig. \ref{fig:Bandstruktur}(b), electrons are driven into the UV QW ($\alpha^+_{n}$) whereas holes are hindered by the polarization fields ($\alpha^-_{p}$).
This leads to the specific forms of $\beta$ for every structure given below the bandstructures from Fig. \ref{fig:Bandstruktur}. 
\begin{figure}[ht]
\includegraphics[width=1\linewidth]{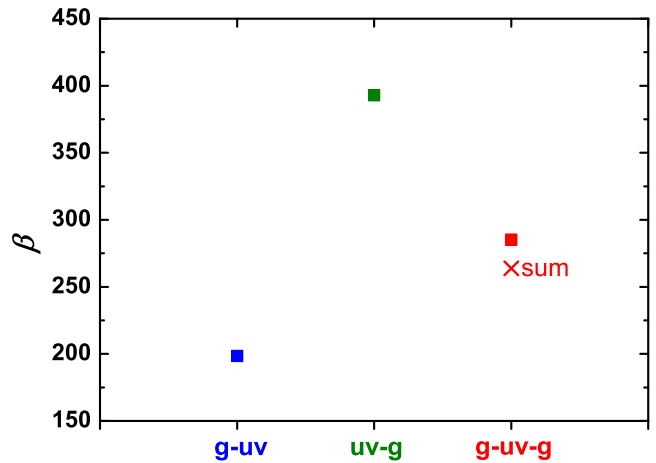}
\caption{\label{fig:Polarisationsfelder} Dependence of $\beta$, and thus the capture probability, on the arrangement of green and UV wells regarding polarization fields as discussed in section \ref{ssec: Pol} (squares). The calculated $\beta^{sum}_{g-uv-g}$ (cross) from the two $\beta$ values of structures uv-g and g-uv using Eq. (\ref{eq: def_sum_g-uv-g}) is in good agreement with the experimental value of structure g-uv-g.}
\end{figure}
Before conducting the intensity-dependent $450\enm$ excitation where UV photons are only generated by Auger processes in the green QW, it was checked by temperature-dependent PL measurements that at $12\eK$ the wells are negligibly affected by SRH recombination.
The experimental $\beta$ values of the samples were calculated according to Eq. (\ref{eq: calc}) and are plotted in Fig. \ref{fig:Polarisationsfelder}. It can be clearly seen that $\beta_{g-uv}<\beta_{uv-g}$. This finding contradicts Eq. (\ref{eq: PolNoInfluence}) and one can thus conclude that the polarization fields must have an influence and carriers are more likely to be captured if accelerated by the fields towards the detection well, i.e. 
\begin{equation}
\alpha^+_{n}>\alpha^-_{n}  \text{ and }  \alpha^+_{p}>\alpha^-_{p}. 
\label{eq: Ann}
\end{equation}
Using the equations below the bandstructures of Figs. \ref{fig:Bandstruktur}(a) and \ref{fig:Bandstruktur}(b), the observed inequality of the $\beta$ values results in
\begin{equation}
\text{min}[\alpha^-_{n}\cdot C_{nnp}, \alpha^+_{p}\cdot C_{npp}]>\text{min}[ \alpha^+_{n}\cdot C_{nnp}, \alpha^-_{p}\cdot C_{npp}].
\label{eq: Pol}
\end{equation}
There are four possible cases leading to Eq. (\ref{eq: Pol}).  
In the following, the most reasonable scenarios concerning the lower type of generation rate in the UV well are analyzed. 

It follows from Eq. (\ref{eq: Pol}) that electrons cannot be the limiting recombination partner in both structures since this would contradict Eq. (\ref{eq: Ann}). 
The three other cases will be discussed through the experimental study of the structure g-uv-g shown in \ref{fig:Bandstruktur}(c). 
In this sample, two sources of Auger carriers are present, i.e. a green QW both below and above the single UV well. 
Hence, the UV well can capture both carrier types from both sides once with a higher and once with a lower capture probability depending on the direction of polarization fields. Furthermore, since every carrier in the UV QW can be ascribed to an Auger process in a single green QW, the total number of high-energy carriers in the structure g-uv-g can be explained by the generation rates of both green QWs. 
From the structures g-uv and uv-g it is known how many hot electron-hole pairs are generated from either the bottom or the top grown green QW. Adding the corresponding UV luminescence should result in the number of UV photons in the structure g-uv-g, if no additional or fewer electron-hole pairs arise due to cross-over interaction of the two sources of Auger carriers filling one single capture well. 
To ensure that the number of UV photons is not affected by different excitation densities of the green QWs and thus Auger processes in the individual structures, we add the inverse $\beta$ values instead of the measured UV photons of structures g-uv and uv-g.  The reciprocal value then gives the proportionality factor of the sum of the UV photons normalized to one green QW
\begin{equation} 
\beta^{sum}_{g-uv-g}=2\cdot \left(\frac{1}{\beta_{g-uv}}+\frac{1}{\beta_{uv-g}}\right)^{-1}. \label{eq: def_sum_g-uv-g}
\end{equation} 
Hence, to be comparable to structure g-uv-g, the factor $2$ accounts for the second source of Auger carriers, which enhances the total droop rate by the same factor as can be seen from Eq. (\ref{eq: calc}). 
$\beta^{sum}_{g-uv-g}$ gives thus the proportionality factor of the structure g-uv-g without additional or fewer UV photons due to cross-interactions between the pairs g-uv and uv-g. 
From the comparison of the experimental values reported in Fig. \ref{fig:Polarisationsfelder} we conclude that $\beta^{sum}_{g-uv-g}=\beta_{g-uv-g}$
and thus
\begin{gather} 
\text{min}[\alpha_{n}^-\cdot C_{nnp},\alpha_{p}^+\cdot C_{npp}]+\text{min}[\alpha_{n}^+\cdot C_{nnp},\alpha_{p}^-\cdot C_{npp}]  \nonumber \\
= \text{min}[(\alpha_{n}^-+\alpha_{n}^+)\cdot C_{nnp},(\alpha_{p}^-+\alpha_{p}^+)\cdot C_{npp}].  \label{eq: sum_g-uv-g}            
\end{gather} 
If holes are limiting in all structures as proposed before, Eq. (\ref{eq: sum_g-uv-g}) is satisfied. In this case, the UV photons are a measure of the hot hole generation rates in the three structures. Adding the number of holes from both sides of the UV QW yields thus the number of holes in the structure g-uv-g. 

This consideration can be used to exclude the possible scenario in which holes limit in the structure uv-g and electrons in the structure g-uv. This would happen if the generation rate in the UV wells is always higher for the type of high energy carrier which is driven into the well, i.e. $\alpha^-_{n}\cdot C_{nnp}<\alpha^+_{p}\cdot C_{npp}$ and $\alpha^+_{n}\cdot C_{nnp}>\alpha^-_{p}\cdot C_{npp}$. 
In this case, the structure g-uv-g would exhibit a higher generation rate of hot electron-hole-pairs, and thus UV luminescence, than if the UV photons of structures g-uv and uv-g are added. The reason for this is that hot holes lacking a recombination partner from the bottom recombine with excess electrons originating from the top grown green well. This interaction, which would lead to $\beta^{sum}_{g-uv-g}>\beta_{g-uv-g}$, is not observed experimentally. 

In the fourth case of Eq. (\ref{eq: Pol}), the limiting recombination rate would always be the one where the high energy carriers are driven into the UV wells by the polarization fields, i.e.  $\alpha^-_{n}\cdot C_{nnp}>\alpha^+_{p}\cdot C_{npp}$ and $\alpha^+_{n}\cdot C_{nnp}<\alpha^-_{p}\cdot C_{npp}$. This assumption is not only counter-intuitive, but also contradicts Eq. (\ref{eq: Ann}) in combination with Eq. (\ref{eq: sum_g-uv-g}).

Summarizing these considerations, holes are most likely limiting the UV luminescence in all samples discussed in this section. Besides the evaluation shown here, a second argument for the lower generation rate of holes in the UV wells can be found from literature: It was shown experimentally by our group \cite{Galler2013}, as well as theoretically by Bertazzi et al. \cite{BertazziIWN} for direct processes without phonon interactions, that the Auger coefficient for electrons $C_{nnp}$ exceeds the one for holes $C_{npp}$. Consequently, hot holes are most probably less generated than hot electrons by Auger processes in the green QWs.
Hence, dividing the expression for $\beta_{g-uv}$ by $\beta_{uv-g}$ and utilizing the fact that holes limit the UV luminescence in both structures it follows for $7\enm$ barrier thickness
\begin{equation}
\alpha_{p}^-=0.5 \cdot \alpha_{p}^+
\label{eq: alpha_asym}
\end{equation}
by comparison with the experimental values reported in Fig. \ref{fig:Polarisationsfelder}.

\subsection{Variation of Capture Distance}
\label{ssec: Dis}
\begin{figure}[ht]
 \centering
\includegraphics[width=1\linewidth]{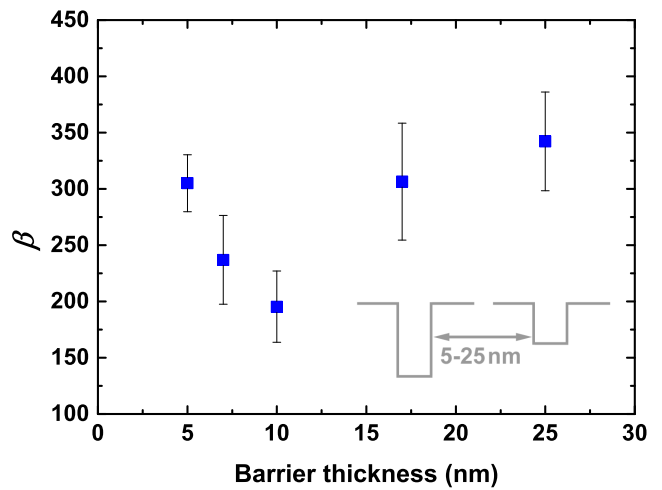}
\caption{\label{fig:Abstandsserie} Change of $\beta$, and thus the capture probability of high-energy carriers generated due to Auger recombination in a green QW, by increasing the distance to a capture well as discussed in section \ref{ssec: Dis}.}
\end{figure}

In a next step, the influence of the distance between generation and detection wells on the capture probability is investigated. For this purpose, the thickness of the AlGaN barrier between the green and UV QW is varied. Since the capture probability of the holes is roughly twice as high in the direction towards the surface, the green QW is grown first in the test structures, i.e. sample g-uv shown in Fig. \ref{fig:Bandstruktur}(a) is the reference. The results from the evaluation of $\beta$ are shown in Fig. \ref{fig:Abstandsserie}. 
As can be seen, the capture probability decreases both towards thin and thick barriers and an optimal width of about $10\enm$ is observed. The reason for the higher $\beta$ of the structures featuring smaller distances has two origins: First, by decreasing the barrier thickness, the capture volumes for high-energy carriers of the green and the UV well start to overlap. Consequently, some of the hot electrons or holes which were scattered into the capture volume of the UV well will be drawn off by the green QW, reducing the number of generated UV photons. Whose capture volume is more effective depends on several mechanisms such as the drift-diffusion component to the wells and the coupling of unbound to bound states in each QW.
Second, due to the stronger coupling of the two wells, the probability for quantum tunneling from the carriers initially captured in the UV back into the energetically more favorable green QW increases. On the other hand, using thick barriers, the high-energy carriers are more likely to relax via phonons and recombine non-radiatively before they can be captured by the UV QW. Such a loss of average energy was also predicted by Monte Carlo simulations of hot-electron transport. \cite{Sadi2014} Nevertheless, a significant fraction of Auger-generated electrons and holes move at least $25\enm$ before recombining non-radiatively. Applying an electric field which additionally accelerates the carriers can lead to the detection of high-energy electrons as far as $200\enm$ from the generating well\cite{Iveland2013}, although with such a low efficiency that no significant UV signal would be obtained in our approach. 

Depending on the variation of $\alpha_{n}^-$ and $\alpha_{p}^+$ with barrier thickness, the trend in Fig. \ref{fig:Abstandsserie} might be dominated by electrons and/or holes which limit the detection efficiency.
As discussed in section \ref{ssec: Pol} for the structures with $7\enm$ barriers, holes most probably limit the recombination in the UV wells. For larger distances, the increase in $\beta$ is likely due to a decrease in $\alpha_{p}^+$ since holes move much slower in real space than electrons and thus have more time to relax to the bandedge and recombine non-radiatively. 
Consequently, holes limit the detection efficiency for all distances in Fig. \ref{fig:Abstandsserie} besides the $5\enm$ case, no matter if relaxation or tunneling is the prevailing hot carrier loss.
Hence, a change in type of carrier limiting the UV emission seems unrealistic in the investigated thickness regime and the trend of $\beta$ shown in Fig. \ref{fig:Abstandsserie} is expected to be dominated by $\alpha_{p}^+$.

\subsection{Interaction of Capture Volumes}
\label{ssec: Int}
\begin{figure}[ht]
\includegraphics[width=0.8\linewidth]{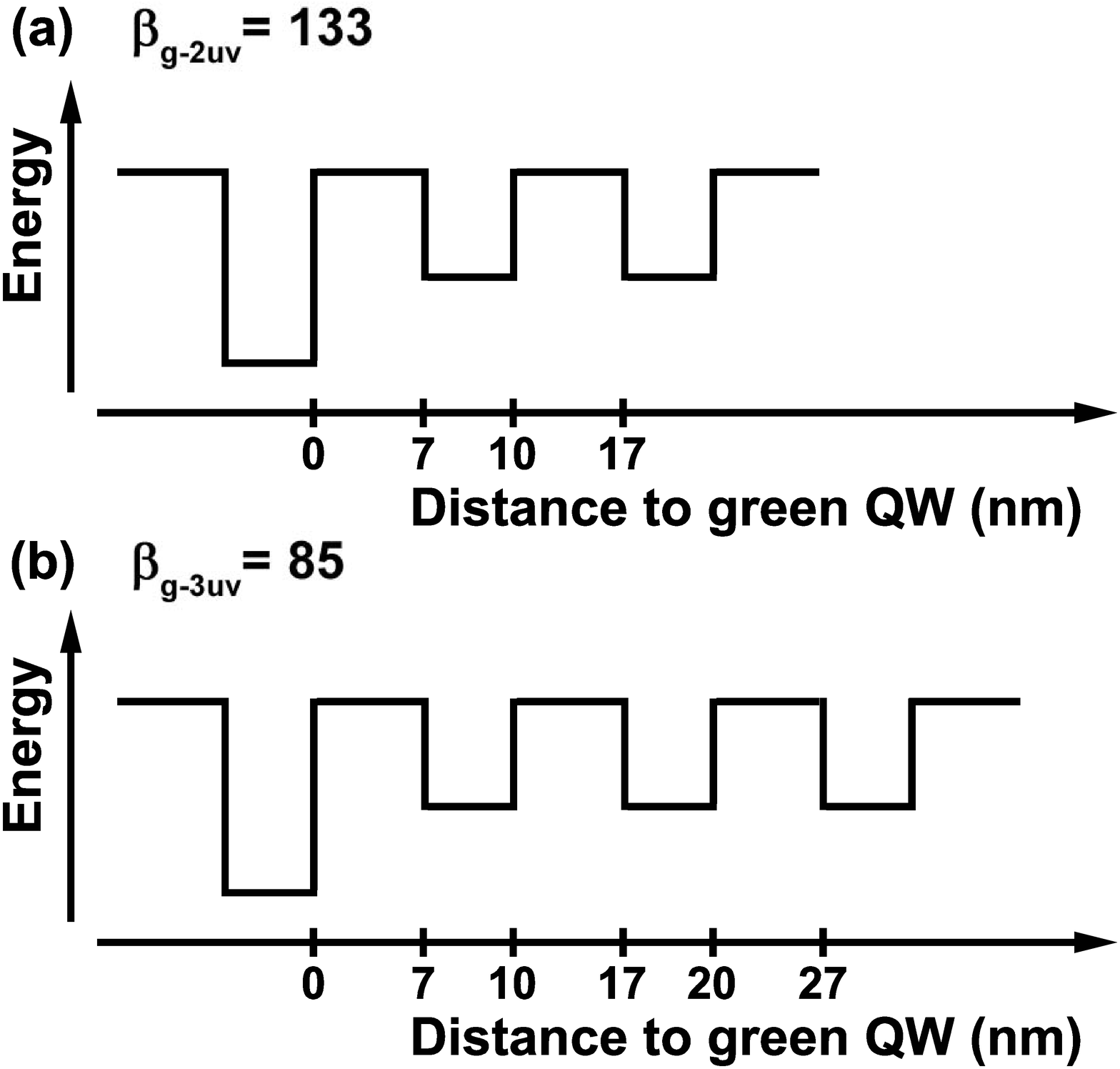}
\caption{\label{fig:virtualBeta} Schematic illustration of the material conduction band of the samples from section \ref{ssec: Int} with (a) two and (b) three UV wells grown after a single green QW. The experimental $\beta$ values determined for both structures are indicated.}
\end{figure}

The next question which arises is the following: 
Is it possible to improve the detection efficiency by adding up the number of UV photons generated at different distances from the green QW, i.e. by using a structure with more than one capture well?
To answer this, two samples were grown with either two or three UV QWs grown above a single green QW. 
The Al$_{0.39}$Ga$_{0.61}$N barriers between every $3\enm$ thick well were chosen to be $7\enm$ thick. The second and the third UV well are thus located $17\enm$ and $27\enm$ behind the green QW as indicated in Fig. \ref{fig:virtualBeta}.  

The $\beta$ values for these positions without the other UV QWs have been determined in section \ref{ssec: Dis}. Based on the considerations leading to Eq. (\ref{eq: def_sum_g-uv-g}), adding the UV photons in the form of $\beta^{-1}$ of these single UV well structures delivers $\left(\beta^{sum}\right)^{-1}$ of a structure including all the UV wells at the same distances to the green QWs as the single UV well structures. The obtained  $\beta^{sum}$ will only have the same value as the experimental $\beta$ of the equivalent structure if there is no interaction between the generation of hot electron-hole pairs in each UV well. 
For instance, $\beta^{sum}_{g-2uv}$ of the sample with two QWs grown after the green QW can be calculated from the two $\beta$ values of section \ref{ssec: Dis} based on $7\enm$ and $17\enm$ barrier thickness
\begin{equation}
\beta^{sum}_{g-2uv}=\left(\frac{1}{\beta_{g-uv}(7\enm)}+\frac{1}{\beta_{g-uv}(17\enm)}\right)^{-1}=131.
\end{equation}
The third QW can be approximated with the $\beta$ of the $25\enm$ case, leading to $\beta^{sum}_{g-3uv}=95$.
Both values are very close to the experimentally determined values $\beta_{g-2uv}=133$ and $\beta_{g-2uv}=85$ for the structures shown in Fig. \ref{fig:virtualBeta}.
Therefore, we conclude that the trajectory of high-energy carriers is hardly affected by inserting UV QWs, i.e. the capture probability of a single UV well is independent from the presence of other capture wells within the investigated samples. In addition, the fields in the QWs, which result from interface charges, have a negligible influence on the transport of the hot carriers. This comes as no surprise as only $3\enm$ of inverse field within a $17\enm$ barrier will not significantly change the capture probability in contrast to changing the total field of a $7\enm$ thick barrier which led to a capture probability reduced by $50\eProzent$ (cp. Eq. (\ref{eq: alpha_asym})). Consequently, every UV QW has a finite capture volume which extends less than $7\enm$ in growth direction and a significant fraction of the Auger-generated electrons and holes are not captured by the UV QWs next to the green QW but are able to traverse them.

\subsection{Increase of Capture Volume}
\label{ssec: Vol}
\begin{figure}[ht]
 \centering
  \includegraphics [width = 1\linewidth ] {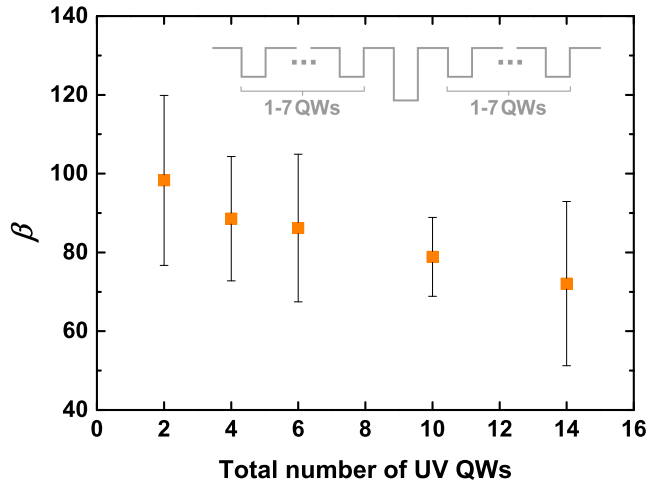} 
\caption{\label{fig:Anzahlserie} Dependence of $\beta$ on the number of UV QWs which correlates with the capture volume for high-energy carriers (section \ref{ssec: Vol}).}
\end{figure}

As suggested from the findings of the previous section, the detection efficiency can be improved by increasing the capture volume. Although the capture probability is lower in the direction towards the substrate as discussed for Eq. (\ref{eq: alpha_asym}), adding up the UV QWs on both sides of the green QW is expected to increase the total capture rate. 
As a last series, the investigated samples feature an increasing number of UV QWs on both sides of one green QW, each separated by $7\enm$ thick AlGaN barriers as schematically shown in the inset of Fig. \ref{fig:Anzahlserie} where the experimentally determined $\beta$ values are summarized. 

By increasing the number of QWs more high-energy carriers are captured and $\beta$ is reduced. 
Using seven UV wells on each side of the green QW, at least $2.8\eProzent$ of the droop can be ascribed to Auger recombination without accounting for the reduced capture probability in the wells below the green well due to the polarization fields or the asymmetry in type of Auger coefficients. Since the decrease of $\beta$ does not yet show a saturation, further improvement in detection efficiency due to the increase of capture volume seems possible.

\section{conclusion}
In summary, the influence of three factors on the capture probability of high-energy carriers generated by Auger recombination into a detection well was investigated. First, polarization fields lead to an enhanced capture probability in the preferred transport direction which is opposite for hot electrons and holes. Second, by increasing the barrier between the generating and capture well of high-energy carriers, an optimum distance for the capture probability is obtained due to the strong capture of the energetically more favorable green QW for thin and carrier relaxation for thick barriers. 
Third, every UV QW has a finite capture volume and thus increasing the number of wells improves the total capture probability. With this design variation the detection efficiency of Auger recombination contributing to droop could be increased from the previous value of $1\eProzent$ \cite{Binder2013} to $2.8\eProzent$ and further improvement seems possible, since no saturation in the enhancement of capture probability with increasing capture volume has been found yet.
The investigations further support the previous finding \cite{Galler2013} that electron-electron-hole exceed electron-hole-hole Auger processes in InGaN QWs if the densities of both carrier types are similar. This asymmetry adds to the limitations on capture probability to decrease the possibility of visualizing Auger processes contributing to droop in photoluminescence. For this reason, the obtained detection efficiency as a lower limit largely underestimates the real impact of Auger recombination on the reduction of efficiency in (AlInGa)N based LEDs. 

\section*{acknowledgment}
We gratefully acknowledge the financial support of the European Union FP$7$, NEWLED project, grant number $318388$.

\end{document}